\documentclass{article}

\usepackage{geometry}
\usepackage{multibib}
\newcites{method}{Methods References}
\usepackage[utf8]{inputenc}
\usepackage{amsmath}
\usepackage{amssymb,amsfonts,latexsym}
\usepackage{bm}
\usepackage[mathcal]{euscript}
\usepackage{graphicx}
\usepackage{epsfig}
\usepackage{color}
\usepackage{pifont,wasysym,marvosym}
\usepackage{textcomp}
\usepackage{comment}
\usepackage{epstopdf}
\usepackage{gensymb}
\usepackage{dcolumn}
\usepackage{hyperref}
\usepackage{physics}
\usepackage{lineno}

\usepackage{authblk}

\usepackage{lipsum} % for dummy text
\begin{document}
\pdfoutput=1

\title{\textbf{Tuning the buckling sequences of metamaterials using plasticity}}

\author[1]{Wenfeng Liu}
\author[2]{Bernard Ennis}
\author[1,$\dagger$]{Corentin Coulais}

\affil[1]{\small{Institute of Physics, Universiteit van Amsterdam, Science Park 904, 1098 XH Amsterdam, The Netherlands}}
\affil[2]{Tata Steel Nederland, 1970 CA IJmuiden, The Netherlands}
\affil[$\dagger$]{Corresponding Author: coulais@uva.nl}
\date{}
\maketitle

{\bf
Material nonlinearities such as hyperelasticity, viscoelasticity, and plasticity have recently emerged as design paradigms for metamaterials based on buckling. These metamaterials exhibit properties such as shape morphing, transition waves, and sequential deformation. 
In particular, plasticity has been used in the design of sequential metamaterials which combine high stiffness, strength, and dissipation at low density and produce superior shock absorbing performances. 
However, the use of plasticity for tuning buckling sequences in metamaterials remains largely unexplored. 
In this work, we introduce yield area, yield criterion, and loading history as new design tools of plasticity in tuning the buckling load and sequence in metamaterials. We numerically and experimentally demonstrate a controllable buckling sequence in different metamaterial architectures with the above three strategies. 

Our findings enrich the toolbox of plasticity in the design of metamaterials with more controllable sequential deformations and leverage plasticity to broader applications in multifunctional metamaterials, high-performance soft robotics, and mechanical self-assembly.
}

echanical metamaterials using Euler buckling or snap-through buckling exhibit striking properties such as negative Poisson's ratio~\cite{bertoldi2010negative, overvelde2014relating, shim2012buckling}, nonreciprocity~\cite{coulais2017static}, shape morphing~\cite{xia2019electrochemically, siefert2019bio}, and sequential deformations~\cite{coulais2018multi, meng2020multi, florijn2014programmable, shan2015multistable, restrepo2015phase, frenzel2016tailored, chen2021reusable}. 
%Sequential deformations in metamaterials which occur in a well-delineated sequence of buckling steps has been one of the distinctive functionalities of metamaterials. 
Sequential deformations have emerged as a particularly promising mechanism for a wide range of applications, from shock absorption~\cite{shan2015multistable, restrepo2015phase, frenzel2016tailored, chen2021reusable, Liu2024harness} and vibration damping~\cite{dykstra2023buckling,liang2022phase} to soft robotics~\cite{djellouli2024shell, kamp2024reprogrammable, janbaz2024diffusive, melancon2022inflatable} and mechanical computing~\cite{guo2023non, kwakernaak2023counting, bense2021complex, he2024programmable}. 
%Sequential deformations have mainly been achieved in metamaterials whose unit cells in essence follow the snap-through buckling of von Mises trusses~\cite{florijn2014programmable, shan2015multistable, restrepo2015phase, frenzel2016tailored}.% or the buckling of wide beams~\cite{coulais2018multi, chen2021reusable}.
The mechanism of the sequential deformations in metamaterials is two-fold. First, there is a weakening mechanism that leads to localisation of the deformation. In metamaterials, this mechanism is typically realized using buckling with a negative stiffness, such that the metamaterial locally buckles or snaps---since the stiffness is negative from the onset of instability, it is more energetically favourable to increase further deformations locally. The negative stiffness can be caused by Euler buckling in combination with a material's hyperelasticity~\cite{coulais2015discontinuous, lubbers2017nonlinear, chen2021reusable, Liu2024harness}, a snap-through mechanism that in essence 
follows the snap-through buckling of a von Mises truss
~\cite{florijn2014programmable,shan2015multistable, restrepo2015phase, frenzel2016tailored, nadkarni2016unidirectional}, or the balloon instability~\cite{overvelde2015amplifying}. Second, once the localized deformation has taken place, it is superseded by a stiffening mechanism. The stiffening mechanism can be contact interactions~\cite{shan2015multistable, restrepo2015phase,overvelde2015amplifying, Liu2024harness}, stretching~\cite{florijn2014programmable,rafsanjani2016snapping,frenzel2016tailored,rafsanjani2019propagation} , fluid incompressiblity~\cite{vasios2021universally, martinez2024fluidic, djellouli2024shell}, or magnetic repulsion~\cite{nadkarni2016unidirectional, veenstra2024non}.  
It is possible to tune the buckling sequence in these metamaterials by controlling either or both of these softening and stiffening mechanisms~\cite{florijn2014programmable, frenzel2016tailored, coulais2018multi}, but the options to do so are typically limited to geometrical design. In this work, we show that plasticity offers additional design freedom to tune sequential buckling.

%Thus, these metamaterials are inherently compliant because of the bending-dominated architectures or elastomeric consecutive materials, limiting the use of sequential metamaterials in high-tech applications which need load bearing and high energy dissipation. 

Plasticity has been recently introduced as a new route to create metamaterials that can buckle in an arbitrary large sequence of steps~\cite{Liu2024harness}. Specifically, a regime of elastoplastic buckling, wherein the stress distribution across the buckling ligament is highly asymmetric---one part is compressed into the plastic regime whereas the other is unloaded elastically---has been shown to lead to a negative stiffness right at the onset of buckling. Such ``yield buckling'' leads to a well-defined sequence of buckling steps. In contrast to the sequential metamaterials
based on Euler buckling of hyperelastic materials or on snap-through buckling, sequential metamaterials based on yield buckling can combine low density, high stiffness, high strength, and high dissipation~\cite{Liu2024harness}. What we show here is that yield buckling also provides new ways of tuning the buckling sequence using the yield area, the yield criterion, and the loading history. By the strategic use of plasticity, our work hence opens up the design space for sequential metamaterials.

In this work, we analytically, numerically, and experimentally study the tunability of yield buckling and the buckling sequence in metamaterials consisting of rigid parts connected by ligaments. 
We organize the manuscript into six sections. In section II, we introduce the methods of designs and fabrications, experiments, and numerical simulations. In section III, we first review the regime of yield buckling and explain how it differs from elastic and plastic buckling. We then introduce the yielding area determined by the ligament thickness as the first design tool in tuning the yield buckling load and the yield buckling sequences in metamaterials. In section IV, we introduce the yield criterion adjusted by the ligament orientation as the second tool to tune the yield buckling load and the buckling sequences. In section V, we explore the possibility of tuning the buckling sequence by loading history. We introduce the Bauschinger effect from the kinematic work hardening rule to yield buckling and achieve multiple buckling sequences with loading history. In section VI, we discuss the potential challenges and chances of using plasticity in the design of sequential metamaterials and conclude the manuscript.

%Our work introduces the yield area, yield criteria, and work hardening rules into yield buckling and the design of sequential metamaterials with controllable buckling sequences. Our findings enrich the design tool of plasticity in metamaterials and elevate sequential metamaterials into applications with needs for specific buckling sequences such as multifunctional metamaterials, soft robotics, and mechanical self-assembly.

\section{Methods}

\subsection{Design and fabrication methods}
\begin{figure*}[t!]
\centering
\includegraphics[width=1.0\linewidth]{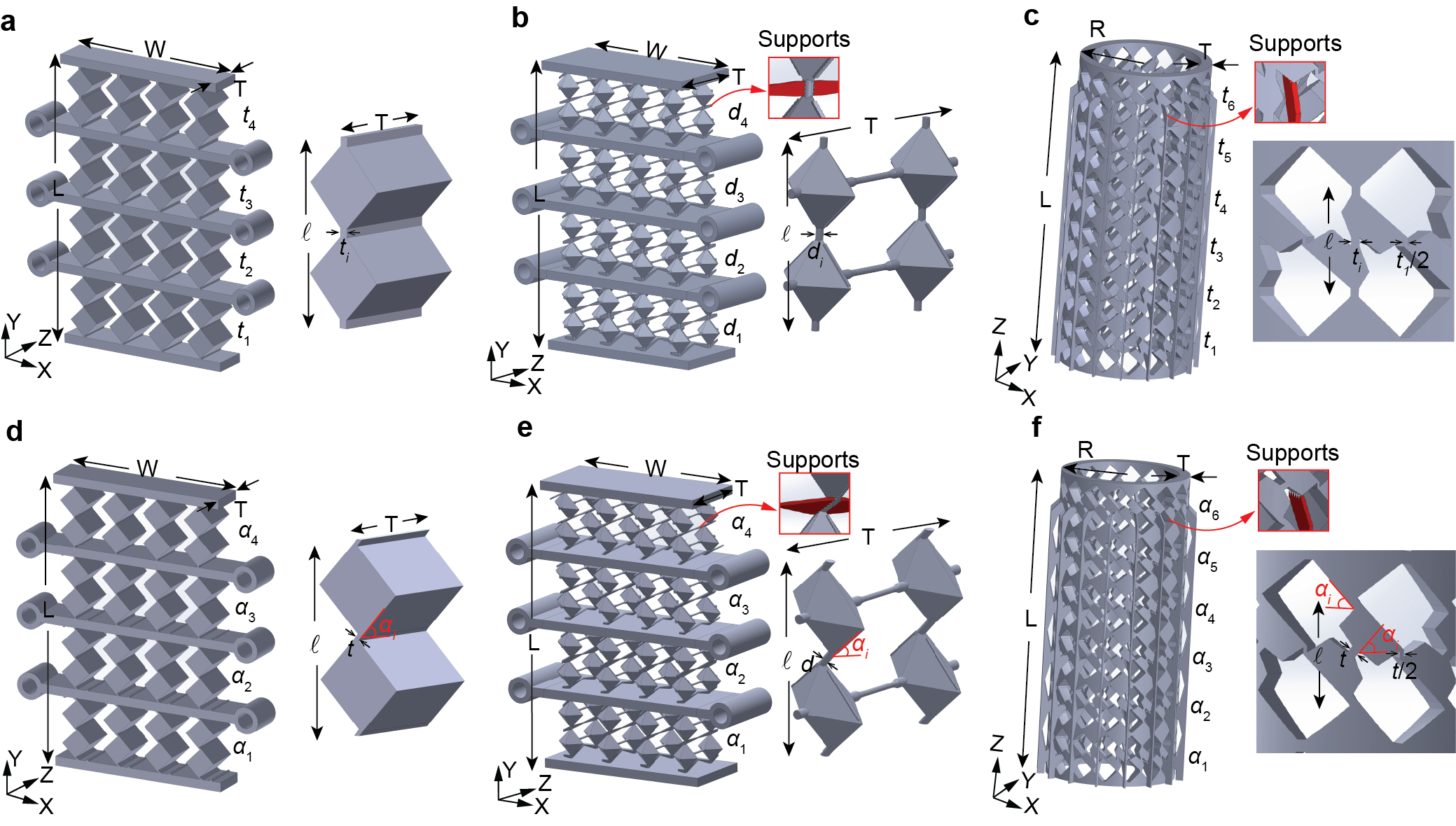}
\caption{\textbf{Geometry design of sequential metamaterials.} \textbf{a and d,} Metamaterials with four layers of rotating-square unit cells. \textbf{b and e,} Metamaterials with four layers of octahedron unit cells. \textbf{c and f,} Metamaterials with six layers of line modes. \textbf{a-c,} The ligament thickness of the metamaterials arithmetic increases from the bottom layer to the top layer. \textbf{d-f,} The ligament orientation of the metamaterials arithmetic reduces from the bottom layer to the top layer.}
\label{fig:1}
\end{figure*}

The sequential metamaterials designed in this work consist of a plurality of buckling unit cells arranged in a periodic pattern, wherein the unit cells comprise rigid parts connected by ligaments that can act as hinges during buckling. Three types of metamaterial architectures are explored here:

\begin{enumerate}
    \item Metamaterials with pairs of rotating squares (Fig.~\ref{fig:1}a and d): 
    The metamaterials of length $L=76$ mm, width $W=46.4$ mm, and depth $T=9$ mm consist of four buckling layers of the same length $\ell=18$ mm and separated rigid plates. Each layer has four independent unit cells and each unit cell comprises a pair of rotating squares connected by ligaments. The ligaments have the same dimension in X and Y directions and the thickness $t_{\text{i}}$ (i = 1, 2, 3 and 4) of the four layers arithmetic increases from the bottom $t_{1}=0.4$ to the top $t_{4}=0.76$ in the design of Fig.~\ref{fig:1}a and the ligament orientation angle $\alpha_{\text{i}}$ (i = 1, 2, 3 and 4) of the layers arithmetic decreases from the bottom $\alpha_{1}=90^{\circ}$ to the top $\alpha_{1}=30^{\circ}$ in the design of Fig.~\ref{fig:1}d with the same thickness $t=0.4$ mm. 
   
    \item Metamaterials with pairs of octahedrons (Fig.~\ref{fig:1}b and e):
    The metamaterials of length $L=76$ mm, width $W=46.4$ mm, and depth $T=19$ mm comprise four buckling layers of the same length $\ell=18$ and separated rigid plates. Each layer has four independently buckling unit cells and each unit cell consists of two pairs of octahedrons connected by ligaments. The ligament diameter of the matematerial in Fig.~\ref{fig:1}b arithmetic increases from the bottom $d_{1}=0.64$ to the top $d_{4}=0.88$ and the ligament orientation angle of the metamaterial in Fig.~\ref{fig:1}e arithmetic decreases from the bottom $\alpha_{1}=90^{\circ}$ to the top $\alpha_{1}=30^{\circ}$ with the same diameter $d=0.64$ mm. 
     
    \item Metamaterials with line modes (Fig.~\ref{fig:1}c and f): 
    The metamaterials of outer radius $R$ = 15 mm, thickness $T$ = 2.5 mm, and length $L$ = 70.5 mm consist of six layers of line modes ~\cite{Liu2024harness}. The ligament thickness of the matematerial in Fig.~\ref{fig:1}c arithmetic increases from the bottom $t_{1}=0.43$ to the top layer $t_{6}=0.64$ and the ligament orientation angle of the metamaterial in Fig.~\ref{fig:1}f arithmetic decreases from the bottom $\alpha_{1}=90^{\circ}$ to the top $\alpha_{1}=30^{\circ}$ with the same thickness $t=0.5$ mm. 
\end{enumerate}

We 3D printed such metamaterials with selective laser melting (SLM) technology (GE additive GlabR) in 316L stainless steel. This printing material has a small ratio $E_{\text{t}}/E = 0.25 \%$ between tangent modulus ($E_{\text{t}}\approx500$  MPa) and elastic modulus ($E\approx200$ GPa), and a relatively high yield stress $\sigma_{\text{y}} = 500$ MPa ~\cite{Liu2024harness}. All the metamaterial samples are printed along the Z direction and the metamaterials with a 2D pattern (Fig.~\ref{fig:1}a and d) can be directly printed on a platform without support structures and can be removed from the platform with a wire-cutting machine after printing. For the metamaterials with octahedrons and line modes, support structures are needed during printing. we set supports at the overhang edges of the ligaments and blocks (Fig.~\ref{fig:1}b,c,e, and f). We then removed the supports manually after printing.  

\subsection{Experimental methods}
The static compression of the metamaterials was processed in a universal testing machine (Instron 5985) equipped with a 300 kN load cell and compression plates, which enabled to impose a compressive displacement with an accuracy of 0.01 mm and to record the force with $0.5\%$ of the reading accuracy. We performed a continuous compressing test in the displacement control with a speed of 0.2 mm/s. For the four-layer metamaterials with separating plates, we performed the compression in a U-shape setup where two rigid constraints of the setup are used to prevent the shearing movement, and six pairs of assembled bearings at each side of the plates are used to guide the vertical movement and reduce the friction.  

The quasi-static tests were recorded using a high-resolution camera (Nikon D780 with 105 mm lens, resolution 1080 px $\times$ 1280 px, and frame rate 30 fps). The white dots are manually drawn on the separating plates and squares to facilitate position tracking. We then used Image-J software to track the position of the white dots of each frame in the video. From the dot's position at each frame, we calculate the strain $u/\ell$ of each buckling layer. 
\subsection{Numerical methods}
We use the ``Standard'' solver in the commercial software Abaqus (2020) for the finite-element simulation.

\textit{Model definition}. We analyze the buckling behaviors of a unit cell and two lumped unit cells in the regimes of elastic, plastic, and yield buckling. The unit cell of length $\ell$ consists of a pair of rotating squares connected by ligaments of thickness $t$ and orientation $\alpha$ (Fig.~\ref{fig:2}a inset and Fig.~\ref{fig:4}a). 
To study the buckling sequence affected by the length $\ell$ and ligament thickness t of the unit cell, we model a structure with two lumped unit cells by varying the length ratio $\ell_{1}/\ell_{2}$ and the ligament thickness ratio $t_{1}/t_{2}$ at constraints of $\ell_{1}+\ell_{2}= 40$ mm and $t_{1}+t_{2}=2$ mm (Fig.~\ref{fig:2}c). 
We then study the effect of ligament orientation on the buckling behaviors by varying ligament orientation from $90^\circ$ to $5^\circ$ in a unit cell of length $\ell$ = 20 mm and thickness $t$ = 1 mm (Fig.~\ref{fig:5}a-c) and then modeling the buckling sequence in a two-layer structure by varying angle ratio $\alpha_{1}/\alpha_{2}$ and length ratio $\ell_{1}/\ell_{2}$ at constraints of $\ell_{1}+\ell_{2}= 40$ and $\alpha_{1}+\alpha_{2}=180^\circ$ (Fig.~\ref{fig:4}b).
To ensure the buckling unit cells always sit in the same buckling regime, we use a linear elastic model of Young's modulus $E = 200$ GPa and Poisson's ratio $\mu=0.3$ for the elastic buckling regime, a bilinear elastoplastic model of $E_t/E=0.25$ for the plastic buckling regime, and of $E_t/E=0.005$ for the yield buckling regime with the yield stress $\sigma_{\text{y}}$ =500 MPa (see Fig.~\ref{fig:2} and  Fig.~\ref{fig:5}).
Finally, we model the loading history effect in a unit cell of length $\ell$ = 20 mm and thickness $t$ = 1 mm (Fig.~\ref{fig:7}d) and in a structure of the two lumped unit cells. We use the bilinear elastoplastic model of $E_t/E=0.1$ and $\sigma_{\text{y}}$ =1750 MPa with kinematic work hardening rule (Fig.~\ref{fig:7}de). We use plane stress conditions with quadratic triangular elements (CPS6) and construct the mesh so that all the ligaments in the unit cells are 10 elements across.

\textit{Boundary conditions}. For the buckling unit cell, we fix the bottom boundary and only allow vertical displacement at the top boundary. For the two-layer structure, we confine the sides of the separated plates and only allow vertical movement.

\textit{Analysis}. We perform two types of analysis: linear eigenmode analysis, where we calculate the lowest eigenmodes; and nonlinear bifurcation analysis, where a displacement imperfection from the linear analysis is introduced into the nonlinear compression (Static step). We make the imperfection proportional to the length of the unit cell $\ell$ and fix it to $0.001$ unless specified otherwise. We track the deflection $w$ of the middle ligament of the unit cell and use the bifurcation point on the deflection curve to catch the onset of buckling.

\section{Tuning sequential buckling by yielding area}
\begin{figure*}[t!]
\centering
\includegraphics[width=1.0\linewidth]{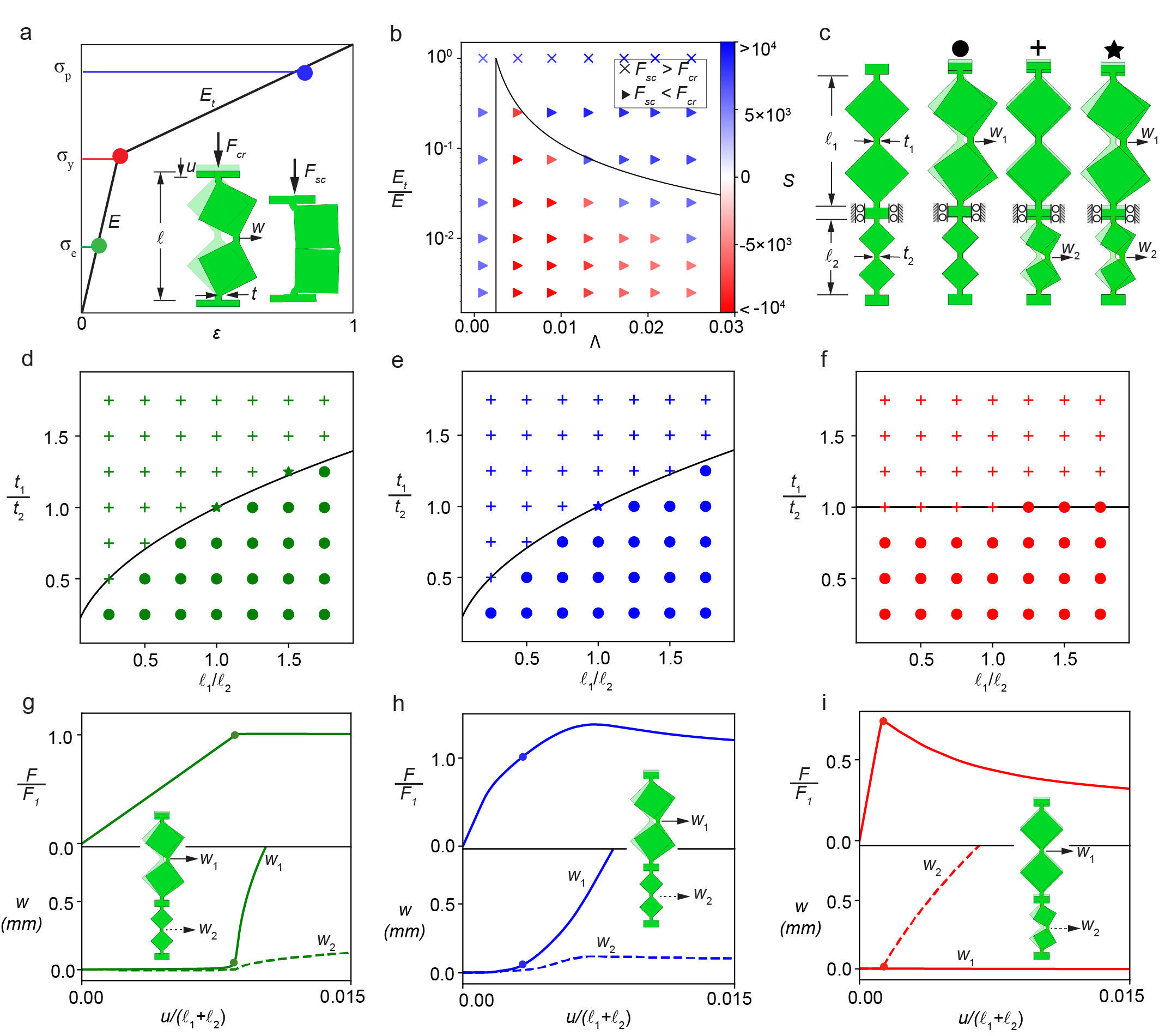}
\caption{\textbf{Yield buckling and buckling sequences.} \textbf{a,} Stress $\sigma$ vs. strain $\varepsilon$ for a bilinear elastoplastic model with Young modulus $E$, yield stress $\sigma_{\text{y}}$ and tangent modulus $E_t$. 
Insets: unit cell with a pair of rotating squares connected by ligaments buckles at a load $F_{\text{cr}}$ and enters contact at a load $F_{sc}$.
The green (before yielding), red (at yielding), and blue (after yielding) markers denote three stress states of the ligaments when the unit starts to buckle. 
\textbf{b,} Post-buckling stiffness S vs. aspect ratio of the unit cell $\Lambda$ and ratio between yield stress and Young’s
modulus $\sigma_{\text{y}}/E$. The red triangles denote the yield buckling regime defined by $S < 0$~\cite{Liu2024harness}.
\textbf{c,} A two-layer strcuture, consisting of two pairs of rotating squares of lengths $\ell_{1}$, $\ell_{2}$ and ligament thicknesses $t_{1}$, $t_{2}$. Three different buckling sequences can be observed: starting from the top layer (circle), bottom layer (plus), or simultaneous(star). \textbf{d-f,} Buckling sequence of the structure in the regimes of elastic (d), plastic (e) and yield buckling (f) vs. length ratio $\ell_{1}/\ell_{2}$ and thickness ratio $t_{1}/t_{2}$ of the two-layer structure. Solid lines are theoretical predictions. Data points are from simulations. \textbf{g-i,} Buckling behaviors of a two-layer structure with $\ell_{1}/\ell_{2}$=0.175 and $t_{1}/t_{2}$=0.125 in the regime of elastic buckling (g), plastic buckling (h), and yield buckling (i). The force $F$ is normalized by the critical buckling load of the top unit cell (top) $F_{1}$. The critical buckling load is determined by the bifurcation point of the deflection $w$ curves (bottom).}
\label{fig:2}
\end{figure*}

We first review the yield buckling behavior in a pair of squares connected by ligaments with an elastoplastic material model (Fig.~\ref{fig:2}a).
Three regimes of buckling---elastic buckling, plastic buckling, and yield buckling---can be observed depending on the aspect ratio $\Lambda$ of the unit cell (here, $\Lambda=t/(2\ell)$) and the yield strain $\sigma_{\text{y}}/E$ or the ratio between tangent modulus and Young's modulus $E_t/E$ of the base material (Fig.~\ref{fig:2}ab). When the aspect ratio $\Lambda$ is smaller than the yield strain $\sigma_{\text{y}}/E$, the unit cell buckles in the elastic regime (Fig.~\ref{fig:2}a green) with a critical load $F^{\text{e}}$,
\begin{equation}\label{elastic buckling load}
F^{\text{e}}=E A \Lambda,%\frac{Et^{2}}{2\ell}.
\end{equation}
where $A$ is the cross-section area of the ligament (here $A=t\times1$ for a unit depth).
When the aspect ratio $\Lambda$ is larger than the yield strain $\sigma_{\text{y}}/E$ and the moduli ratio $E_{\text{t}}/E$ is sufficiently large, the unit cell buckles in the plastic hardening regime (Fig.~\ref{fig:2}a blue). 
The upper bound of plastic buckling is defined by the reduced modulus load
\begin{equation}\label{Plastic buckling load}
F^{\text{p}}=E_r A \Lambda, %\frac{E_{\text{r}}t^{2}}{2\ell}.
\end{equation}
where $ E_{\text{r}}=4EE_{\text{t}} / (\sqrt{E_{\text{t}}}+\sqrt{E})^{2}$ is the reduced modulus~\cite{cedolin2010stability}.
When the aspect ratio $\Lambda$ is larger than the yield strain $\sigma_{\text{y}}/E$ and when the moduli ratio $E_{\text{t}}/E$ is sufficiently small,
then buckling occurs precisely at the yield point (Fig.~\ref{fig:2}a red) with a regime defined by
\begin{equation}
\frac{E_r}{E}\Lambda    <\frac{\sigma_{\text{y}}}{E}< \Lambda .
\label{eq:snapthroughprediction}
\end{equation}
We verify such prediction in numerical simulations by varying the moduli ratio $E_t/E$ and the aspect ratio $\Lambda$ (Fig.~\ref{fig:2}b see also varying the yield strain $\sigma_{\text{y}}/E$ and the aspect ratio $\Lambda$ in Fig.~\ref{fig:7}c). We find a good agreement by measuring the buckling stiffness and the postbuckling load before self-contact (Fig.~\ref{fig:2}b and Fig.~\ref{fig:7}c red triangles). The results confirm that only yield buckling shows the negative stiffness right at the onset of buckling~\cite{Liu2024harness}.

Now we study the buckling sequence in a two-layer structure with two lumped buckling unit cells of lengths $\ell_{1}, \ell_{2}$ and ligament thicknesses $t_{1}, t_{2}$ (Fig.~\ref{fig:2}c). When the shearing movement of the structure is confined, three possible buckling sequences can be observed in the structure: buckling starts from the top layer (circle), starts from the bottom layer (plus), or simultaneously (star). 

We first study the buckling sequence of the structure in the elastic and plastic buckling regimes. We assume the two unit cells of the structure both sit in the same buckling regime. From the critical buckling loads of the single unit cell (Eqs.~(\ref{elastic buckling load}) and ~(\ref{Plastic buckling load})), we can get the boundary between the two sequences (circle and plus) as:
\begin{equation}\label{Elastic and plastic buckling order}
\frac{F_{1}^{\text{e}}}{F_{2}^{\text{e}}}=\frac{F_{1}^{\text{p}}}{F_{2}^{\text{p}}}=\left(\frac{t_{1}}{t_{2}}\right)^2\frac{\ell_{2}}{\ell{1}},
\end{equation}
where $F_{1}^{\text{e}}$,$F_{2}^{\text{e}}$ and $F_{1}^{\text{p}}$, $F_{2}^{\text{p}}$ are the elastic and plastic buckling loads of the two unit cells.

In the yield buckling regime, the buckling load depends on the yield area of the ligament cross-section $t$ and the yield stress $\sigma_{\text{y}}$. For the same yield condition, the sequence boundary of the two-layer structure can be driven as (Eq.~\ref{Yield buckling order}): 
\begin{equation}\label{Yield buckling order}
\frac{F_{1}^{\text{y}}}{F_{2}^{\text{y}}}=\frac{t_{1}}{t_{2}},
\end{equation}
where $F_{1}^{\text{y}}$, $F_{2}^{\text{y}}$ are the yield loads and $t_{1}$, $t_{2}$ are the ligament thicknesses of the two unit cells.

We numerically verify such predictions in the two-layer structure by varying the length ratio $\ell_{1}/\ell_{2}$ and the thickness ratio $t_{1}/t_{2}$ from 0.25 to 1.75.  We find a good agreement between the simulations and theory (solid lines) in all the elastic, plastic, and yield buckling regimes (Fig.~\ref{fig:2}d-f). In the regimes of elastic and plastic buckling, the buckling sequence of the structure depends on both the ligament thickness ratio $t_{1}/t_{2}$ and the length ratio $\ell_{1}/\ell_{2}$. By contrast, the buckling sequence in the yield buckling regime is dominated by the ligament thickness ratio $t_{1}/t_{2}$. Only when the two unit cells have the same ligament size, the buckling sequence is controlled by the length ratio. Since the negative stiffness at the buckling moment, no simultaneous buckling occurs in the yield buckling regime~\cite{Liu2024harness}. 

Interestingly, in some regimes of yield bucking, the structure shows an opposite buckling sequence compared to the structure in elastic and plastic buckling regimes. We showcase such difference in a structure with thickness ratio $t_{1}/t_{2}=1.25$ and the length ratio $\ell_{1}/\ell_{2}=1.75$ 
where the structure buckles from the top layer in elastic and plastic buckling regimes and buckles from the bottom layer in yield buckling regime (Fig.~\ref{fig:2}g-i). The results indicate that the ligament size difference between the buckling layers is more efficient compared to the length ratio in tuning the buckling sequence and can be used to design sequential metamaterials with a large number of programmable sequences.

\begin{figure*}[t!]
\centering
\includegraphics[width=1.0\linewidth]{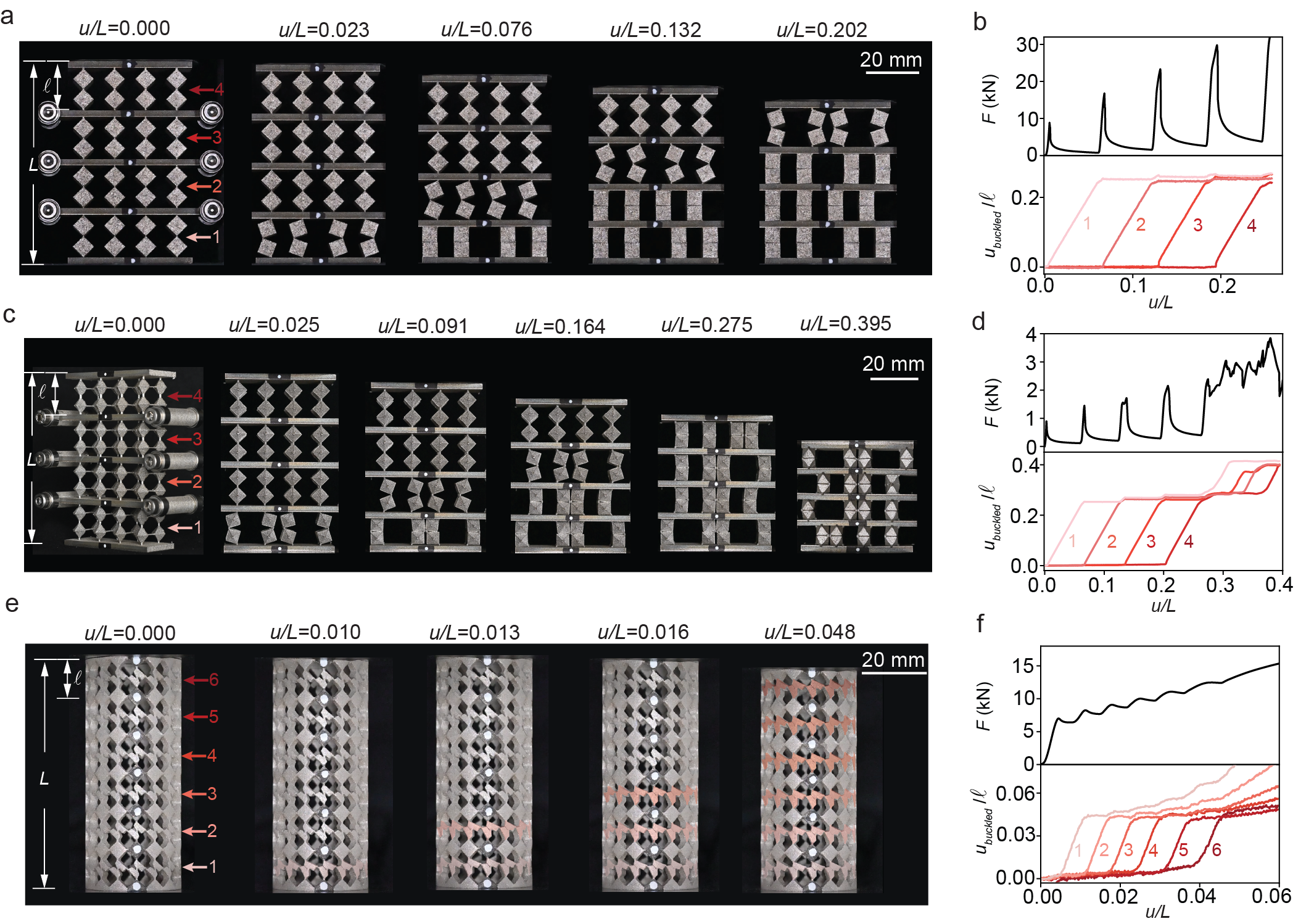}
\caption{\textbf{Sequential yield buckling sequence in metamaterials tuned by the ligament size.} The ligament thickness of each buckling layer in metamaterials gradually increases from the bottom layer to the top layer and the compression speed is 0.2 mm s$^{-1}$ unless indicated otherwise. \textbf{a-d,} Metamaterials with four-layer buckling unit cells and separating plates where the sides of the plates are constrained to avoid shearing and six pairs of load bearings at two sides of the plates are used to guide the movement and reduce friction. \textbf{a and c,} Snapshots of metamaterials with (a) two-square buckling unit cells and (c) with two-octahedron buckling unit cells under uniaxial compression at different strokes $u/L$. \textbf{e,} Snapshots of a metamaterial with six-layer line modes under uniaxial compression at different strokes $u/L$. \textbf{b, d, and f,} Corresponding force $F$ and compressing strain in each buckling layer $u_{\text{buckled}}/\ell$ vs. compressing stroke $u/L$ curves.}
\label{fig:3}
\end{figure*}

Then we experimentally demonstrate programmable buckling sequences in three different types of metamaterials by tuning the ligament thickness (Fig.~\ref{fig:1}a-c). For the three designs of metamaterials, we grudually increase the ligament thickness to program a buckling sequence from the bottom layer to the top layer. 
%In the first metamaterial design, we assemble the unit cells of a pair of squares into a four-layer metamaterial where each layer consists of the same four independent unit cells with length $\ell=18$ mm. The ligament thickness gradually increases from 0.4 mm in the bottom layer to 0.76 mm in the top layer with a difference of 0.12 mm between each adjoined layer. A sample depth is set as T = 9 mm to prevent out-of-plane buckling. To avoid global shearing under compression, we test the sample in a U-shape setup with ball bearings at two sides of separating plates to reduce friction (Fig.~\ref{fig:3}a). 
In the first experiment, we perform the uniaxial static compression on the metamaterial with pairs of rotating squares. The bottom layer of unit cells first buckles at a stroke of $u/L=0.005$ followed by a large force drop in the postbuckling process. The metamaterial then gets stiffened from a self-contact mechanism at a stroke of $u/L=0.065$ and triggers the next step of buckling in the second layer with a higher buckling force. This process repeats until all the layers buckle and enter into contact. As a result, the metamaterial shows four steps of sequential buckling with a sequence from the bottom layer to the top layer (Fig.~\ref{fig:3}a) as designed. The yield buckling load gradually increases in each step (Fig.~\ref{fig:3}b top) and the buckling layer starts to buckle only after the previous layer of squares enters into contact (Fig.~\ref{fig:3}b top). 
In this two-dimensional metamaterial design, each layer only enables a single step of buckling and reaches self-contact and densification at a stroke of $u/L=0.25$. 

A larger compressive stroke and more steps of buckling and steps can be achieved by introducing a three-dimensional buckling unit cell which consists of a pair of octahedrons and connected ligaments at the vertices.
This 3D unit cell enables two steps of buckling under compression. The idea is that the pair of octahedrons will start to rotate in the compressing plane due to the first buckling stability and the two edges of the octahedrons enter into contact such that they make up a new sub-structure. These two octahedrons in the sub-structure will start to rotate along the contacted edges due to the second buckling instability under further compression until two faces of the octahedrons enter into contact. We prove this idea in a metamaterial by replacing the 2D rotating squares in the above design (Fig.~\ref{fig:3}a) with the 3D octahedrons (Fig.~\ref{fig:3}b). 
To prevent out-of-plane buckling and ensure the edge contact after the first step of buckling, we connect two octahedron unit cells in the third dimension.
%In each layer, we arrange four buckling units. The ligament diameter along the loading direction gradually increases from 0.64 mm in the bottom layer to 0.88 mm in the top layer with a difference of 0.08 mm between each adjoined layer.
Under compression, the metamaterial with octahedron unit cells shows two rounds of sequential buckling behaviors. In the first round of buckling, the metamaterial shows the same sequential buckling as the 2D design (Fig.~\ref{fig:3}c) until all the layers get into edge contact at a stroke of $u/L=0.263$. As the compression increases, the reconfigured metamaterial buckles again from the bottom layer to the top layer unitl all the layers enter into face contact at a stroke of $u/L=0.395$. The metamaterial in the second round of buckling has a much less force drop and compressible strokes each step of buckling (Fig.~\ref{fig:3}d). The buckling behaviors in the second round are more complex due to the contact, friction, and non-ideal hinges. Varying ligament thickness also enables a larger difference in the yield buckling load between each buckling layer which can even overcome the global shearing without extra constraints (Fig.~\ref{fig:8}).

In the last design, we vary the ligament thickness in a cylindrical metamaterial with six line modes (Fig.~\ref{fig:1}c). The result of compression shows that the bottom line mode first buckles at a stroke of $u/\ell = 0.005$ and enters into contact at a stroke of  $u/\ell= 0.011$. As the compression increases, the rest of the line modes buckle sequentially until a stroke of $u/\ell= 0.048$. The effective hardening process of the buckling force leads to a controllable buckling sequence from the bottom to the top line mode as designed (Fig.~\ref{fig:3}f). 

%In this design strategy, the ligament thickness of the metamaterials will gradually reduce from the last buckling layer to the first buckling layer, highly increasing the challenge of making 
In the above design strategy, the buckling sequence is dependent upon the ligament thickness; buckling proceeds from the layer with the smallest ligament thickness to that with the largest ligament thickness. This increases the challenge to produce such metamaterials with a larger number of buckling sequences especially at small scale. Can we tune the yield buckling sequence in metamaterials without changing the size of the ligaments? To answer this question we introduce another design tool of plasticity: yield criterion.
%A large number of buckling sequences in metamaterials requires a larger difference in ligament thickness between the first and the last sequence of buckling unit cells, which increases the challenge of manufacture, especially for the metamaterials on a small scale. In the next section, we will introduce another strategy to tune the buckling sequence without changing the thickness of the ligaments.

\section{Tuning sequential buckling by the yield criterion}
\begin{figure*}[b!]
\centering
\includegraphics[width=0.333\linewidth]{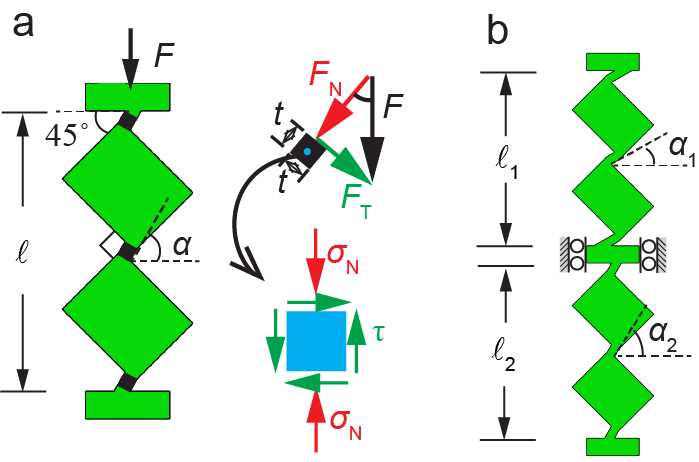}
\caption{\textbf{Buckling structures with orientation angle of ligaments} \textbf{a,} A unit cell consisting of a pair of rotating rectangles and connecting ligaments with an angle $\alpha$ (left). Force and stress analysis of the ligament under loading (right). 
%(\textbf{B, C, D,}) Force $F/F_{\text{cr}}$ and lateral deflection $W$ vs. compressive stroke $u/\ell$ of the unit cell as the ligament angle change in the elastic (B), plastic (C), and yield (D) buckling regimes. The force is normalized by the critical buckling force of the unit cell with $90^{\circ}$ ligament.
\textbf{b,} A two-layer structure consists of two unit cells of lengths $\ell_{1}, \ell_{2}$ and angles $\alpha_{1}, \alpha_{2}$.  
%(\textbf{F, G, H,}) Force $F/F_{\text{cr}}$ and lateral deflection $W$ vs. compressive stroke $u/(\ell_{1}+\ell_{2})$ of the supercell in the elastic (B), plastic (C), and yield (D) buckling regimes. The force is normalized by the critical buckling force of the sub-unit cell with $90^{\circ}$ ligament.
}
\label{fig:4}
\end{figure*} 

\begin{figure*}[b!]
\centering
\includegraphics[width=1.0\linewidth]{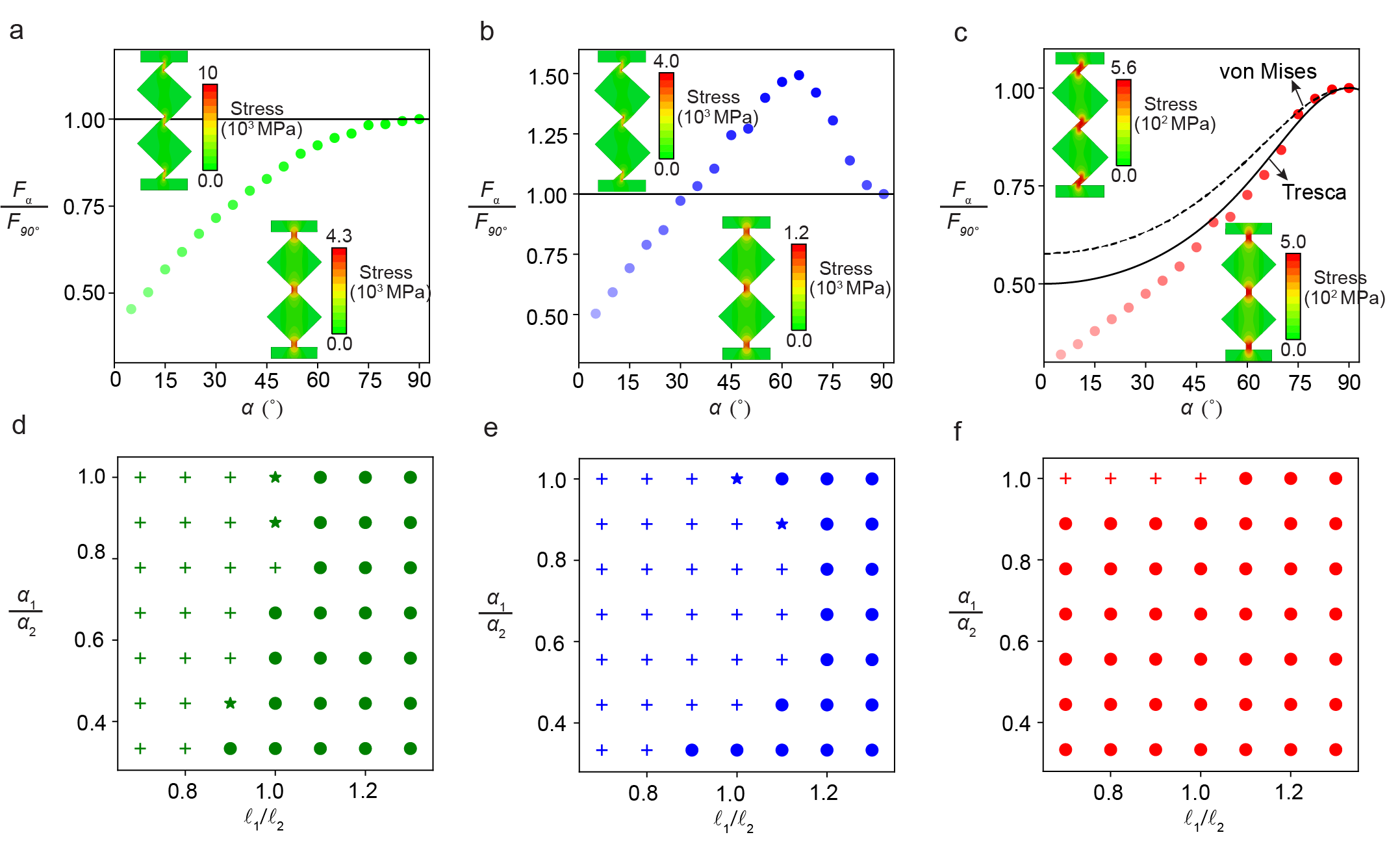}
\caption{\textbf{Yield buckling load and buckling sequence tuned by orientation angle of ligaments.}
\textbf{b-d,} Buckling load $F_{\alpha}/F_{90^\circ}$ vs. orientation angle of ligaments $\alpha$ of a unit cell in the regime of (a) elastic buckling, (b) plastic buckling, and (c) yield buckling. The force $F_{\alpha}$ is normalized by the critical buckling force of the unit cell with $90^{\circ}$ ligament. The insets are the von Mises stress states of unit cells of $45^{\circ}$ and $90^{\circ}$ ligaments at the onset of buckling. 
\textbf{d-f,} The buckling sequence of a two-layer structure in the regime of (d) elastic buckling, (e) plastic buckling, and (f) yield buckling changes with the ratio of orientation angle $\alpha_{1}/\alpha_{2}$ and the length ratio $\ell_{1}/\ell_{2}$ between the two layers unit cells where $\alpha_{2}=90^\circ$ in all the simulations.
}
\label{fig:5}
\end{figure*}
The yield load of the buckling unit cell depends not only on the yield area but also on the yield criterion. When the ligaments of the unit cell are under a uniaxial loading condition, the ligaments yield at a condition of $f(\boldsymbol{\sigma}) =\sigma_{\text{y}}$, where $\sigma_{\text{y}}$ is the yield stress of the based material from a uniaxial stretching test. For a general stress condition $\mathcal{F}(\boldsymbol{\sigma})$, a yield criterion can be described as:
\begin{flalign}\label{general yield Criterion:}
& f(\boldsymbol{\sigma}) = \mathcal{F}(\boldsymbol{\sigma})-k(\xi).
\end{flalign}
We then introduce two main yield criteria for isotropic materials: the von Mises yield criterion and the Tresca yield criterion~\cite{lubliner2008plasticity}.

The von Mises yield criterion suggests that the material yields when the square root of the second stress deviator $\sqrt{J_{2}}=\sqrt{\frac{3}{2}S_{\text{ij}}S_{\text{ij}}}$ achieves a critical value $k(\xi)$: 
\begin{equation}
   f(\sigma)=\sqrt{\frac{3}{2}S_{\text{ij}}S_{\text{ij}}}-k(\xi),
\end{equation}
where $S_{\text{ij}}=\sigma_{\text{ij}}-\frac{1}{3}\sigma_{\text{kk}}\delta_{\text{ij}}$. The second stress deviator $J_{2}$ can be also described by the principle stress $J_{2}=\frac{1}{6}[(\sigma_{1}-\sigma_{2})^{2}+ (\sigma_{1}-\sigma_{3})^{2}+ (\sigma_{2}-\sigma_{3})^{2}]$. and the von Mises yield criterion is:
\begin{equation}\label{Mises Criterion:}
f(\boldsymbol{\sigma})=\sqrt{\frac{1}{6}[(\sigma_{1}-\sigma_{2})^{2}+ (\sigma_{1}-\sigma_{3})^{2}+ (\sigma_{2}-\sigma_{3})^{2}]}-k(\xi).
\end{equation}
The Tresca yield criterion assumes that yielding occurs when the maximum shear stress over all planes attains a critical value, namely, the value of the current yield stress in shear, denoted $k(\xi)$:
\begin{equation}\label{Tresca Criterion:}
f(\boldsymbol{\sigma})=\frac{1}{2}max\{(\sigma_{1}-\sigma_{2}), (\sigma_{1}-\sigma_{3}), (\sigma_{2}-\sigma_{3})\}-k(\xi).
\end{equation}

We use the general yield criteria in the yield buckling unit cell by introducing shearing stress to the ligaments. To do this, we rotate the ligaments with an orientation angle $\alpha$ where the pair of squares change to a pair of rectangles (Fig.~\ref{fig:4}a left). The rotating angle of the rectangles in the postbuckling process is set to $45^\circ$ to have a stable contact.
We then do the force and stress analysis for the ligament with an orientation angle $\alpha$ (Fig.~\ref{fig:4}a right). The stress can decompose to a normal stress and a shear stress:
\begin{equation}\label{stress consition}
\sigma_{N}=\frac{F\cos\alpha}{t}, \tau=\frac{F\sin\alpha}{t}.
\end{equation}
The principal stresses are:
\begin{equation}\label{Eq-11}
\begin{matrix}
\begin{split}
&\sigma_{1}=\frac{1}{2}\sigma_{N}+\sqrt{(\frac{\sigma_{N}}{2})^{2}+\tau^{2}},\\
&\sigma_{2}=0,\\
&\sigma_{3}=\frac{1}{2}\sigma_{N}-\sqrt{(\frac{\sigma_{N}}{2})^{2}+\tau^{2}}.
\end{split}
\end{matrix} 
\end{equation}

We use the von Mises Criterion in a uniaxial stretching test ($\sigma_{1}=\sigma_{1}, \sigma_{2}=\sigma_{3}=0$) and get $k(\xi)=\frac{\sqrt{3}}{3}\sigma_{\text{y}}$,
the von Mises Criterion of the ligament of angle $\alpha$ is:
\begin{equation}\label{Eq-12}
\sqrt{\sigma_{N}^{2}+3\tau^{2}}=\sigma_{\text{y}}.
\end{equation}

The Tresca criterion in a uniaxial stretching test has $k(\xi)=\sigma_{\text{y}}$. Then the Tresca criterion of the ligament of angle $\alpha$ is:
\begin{equation}\label{Eq-13}
\sqrt{(\frac{\sigma_{N}}{2})^{2}+\tau^{2}}=\sigma_{\text{y}}/2.
\end{equation}
With the Eqs.~(\ref{Eq-11}) -~(\ref{Eq-13}), the von Mises yield criterion and the Tresca criterion can be simplified to $f(\boldsymbol{\sigma})=g_\textrm{von Mises}(\alpha)\sigma_{\text{y}}$  and $f(\boldsymbol{\sigma})=g_\textrm{Tresca}(\alpha)\sigma_{\text{y}}$, where $g_\textrm{von Mises}=1/\sqrt{(\cos^2\alpha+3\sin^2\alpha)}$ and $g_\textrm{Tresca}(\alpha)=1/\sqrt{(\cos^2\alpha+4\sin^2\alpha)}$ are yield condition factors. 

\begin{figure*}[b!]
\centering
\includegraphics[width=1.0\linewidth]{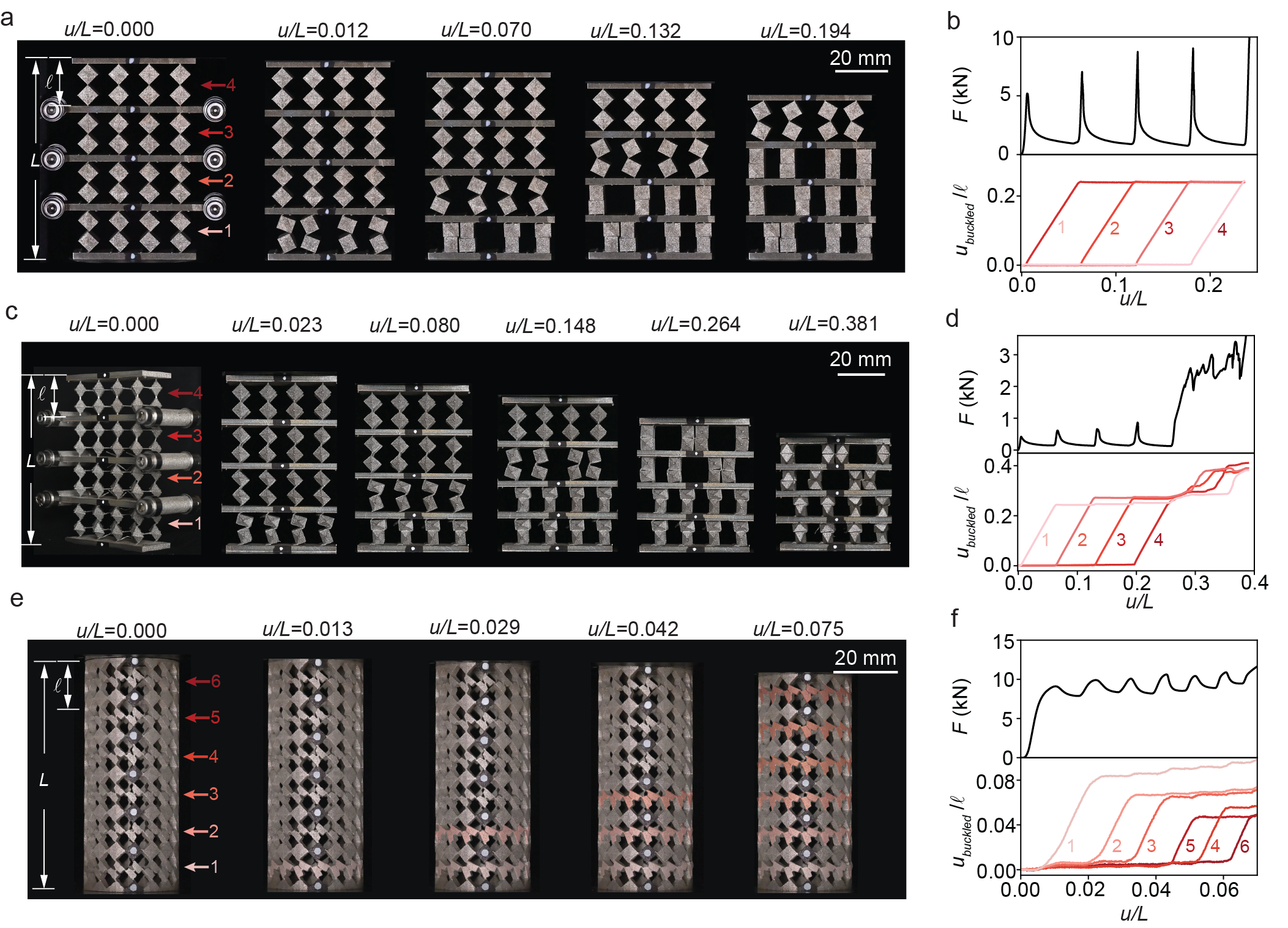}
\caption{\textbf{Sequential yield buckling sequence in metamaterials tuned by the ligament orientation.} The ligament orientation angle of each buckling layer in metamaterials gradually decreases from the bottom layer to the top layer and the compression speed is 0.2 mm s$^{-1}$. \textbf{a-d,} Metamaterials with four-layer buckling unit cells and separating plates where the sides of the plates are constrained to avoid shearing and six pairs of load bearings at two sides of the plates are used to guide the movement and reduce friction. \textbf{a and c,} Snapshots of metamaterials with (a) two-square buckling unit cells and (c) with two-octahedra buckling unit cells under uniaxial compression at different strokes $u/L$. \textbf{e,} Snapshots of a metamaterial with six-layer line modes under uniaxial compression at different strokes $u/L$. \textbf{b, d, and f,} Corresponding force $F$ and compressing strain in each buckling layer $u_{buckled}/\ell$ vs. compressing stroke $u/L$ curves.}
\label{fig:6}
\end{figure*}
We then apply yield criteria to a unit cell with yield buckling and numerically verify it in the unit cell with an orientation angle of the ligaments varying from $\alpha=90^{\circ}$ to $\alpha=5^{\circ}$.
As a comparison, we first explore the angle effect on elastic and plastic buckling. For the case of elastic buckling, the buckling load of the unit cell has a slight change in an angle change from $90^\circ$ to $75^\circ$. When the orientation angle reduces from $75^\circ$, the buckling load starts to decrease rapidly. The reason for the force decrease may be caused by the shearing stress as the shear modulus is smaller than Young's modulus (Fig.~\ref{fig:5}a). For the case of plastic buckling, interestingly, the buckling load first increases as the angle reduces until $60^\circ$ and decreases after that (Fig.~\ref{fig:5}b). 
%The angle orientation in the range of $90^\circ$ to $30^\circ$ shows a strength effect on the buckling load compared to the case of $90^\circ$. 
In the case of yield buckling, the buckling load monotonically decreases as the angle decreases. For a small angle change from $90^\circ$ to $75^\circ$, the von Mises criterion gives a better prediction of the buckling load reduction, and for a larger angle change from $75^\circ$ to $45^\circ$, the load decrease fits better to the Tresca criterion. After the orientation angle is below $45^\circ$, the buckling load linearly decreases as the angle (Fig.~\ref{fig:5}c).

We then study the effect of ligament orientation angle on the buckling sequence of a two-layer structure that consists of two lumped buckling unit cells with the same ligament size. We conduct numerical simulation in a two-layer structure by varying the length ratio $\ell_{1}/\ell_{2}$ from 0.7 to 1.3 and the ligament angle ratio $\alpha_{1}/\alpha_{2}$ from 0.3 to 1.0. The result shows the buckling sequence in the structure is most sensitive to the angle change in the regime of yield buckling compared to elastic and plastic buckling.

In the second experimental demonstration, we tune the yield buckling sequence in the same types of metamaterials by using the ligament orientation. For the 2D and 3D metamaterials with four-layer buckling unit cells, the compressing results show the same buckling sequence from the bottom to the top with the force increase (Fig.~\ref{fig:6}b-d). Since the tunable space with the orientation angle is limited in the range of $90^\circ$, the difference in buckling force between each layer of buckling is much smaller than that of the sequential metamaterials tuned by the ligament thickness (Fig.~\ref{fig:3}b and d). This makes the buckling sequence in metamaterials tuned by the ligament orientation less controllable when the buckling sequence becomes large. In the cylindrical metamaterial with six line-mode buckling layers, we set the orientation angle increasing from 30$^\circ$ at the bottom layer to 90$^\circ$ at the top. The first three buckling steps follow the buckling sequence as designed, yet the fourth and fifth buckling steps do not follow the design. This can be explained by the fact that the buckling loads between these two layers are very close to each other, hence the buckling sequence becomes sensitive to imperfections. Both the strategies above use the combination of plasticity (yield area and yield criterion) and geometry (ligament thickness and orientation) to tune the yield buckling load and sequence in metamaterials. In the next section, we explore the possibility of tuning the yield buckling sequence with the work hardening rule and loading history.

\section{Tuning sequential buckling by loading history}

The yield behavior of elastoplastic materials depends not only on the initial yield criteria but also on the loading history:
\begin{equation}\label{work-haedening}
f(\boldsymbol{\sigma}, \boldsymbol{\rho})=\mathcal{F}(\boldsymbol{\sigma}-\boldsymbol{\rho})-k(\xi),
\end{equation}
where $\boldsymbol{\rho}$ is a parameter related to the loading history.
The von Mises yield criterion can be described as:
\begin{equation}
f(\sigma_{\text{ij}},\rho_{\text{ij}})=\sqrt{\frac{3}{2}(S_{\text{ij}}-\rho_{\text{ij}})(S_{\text{ij}}-\rho_{\text{ij}})}-k(\xi), \quad  S_{\text{ij}}=\sigma_{\text{ij}}-\frac{1}{3}\sigma_{\text{kk}}\delta_{\text{ij}}.
\end{equation}
Two work hardening rules---isotropic and kinematic have been widely used in the plasticity theory. The isotropic hardening simply states that the yield surface expands proportionally in all directions which means the hardening behavior in one direction will not affect the yielding in the opposite direction. The kinematic hardening states that the yield surface translates in one direction as the stress exceeds the initial yield stress
 which means the stress hardening in one direction will lower the yield stress in the opposite direction (also known as the Bauschinger effect)~\cite{lubliner2008plasticity}. 
For the linear kinematic hardening model, $d\rho_{\text{ij}}=\frac{2}{3}E_{\text{t}}d\varepsilon^{\text{p}}_{\text{ij}}$. 
Then the flow law of the elastoplastic material with linear kinematic hardening rule is:
\begin{equation}
    d\varepsilon^{\text{p}}_{\text{ij}}=d\Bar{\varepsilon}^{\text{p}}\frac{3}{2}\frac{S_{\text{ij}}-\rho_{\text{ij}}}{k({\xi})},
\end{equation}
where $\Bar{\varepsilon}^{\text{p}}$ is the accumulated plastic strain.

\begin{figure*}[t!]
\centering
\includegraphics[width=1.0\linewidth]{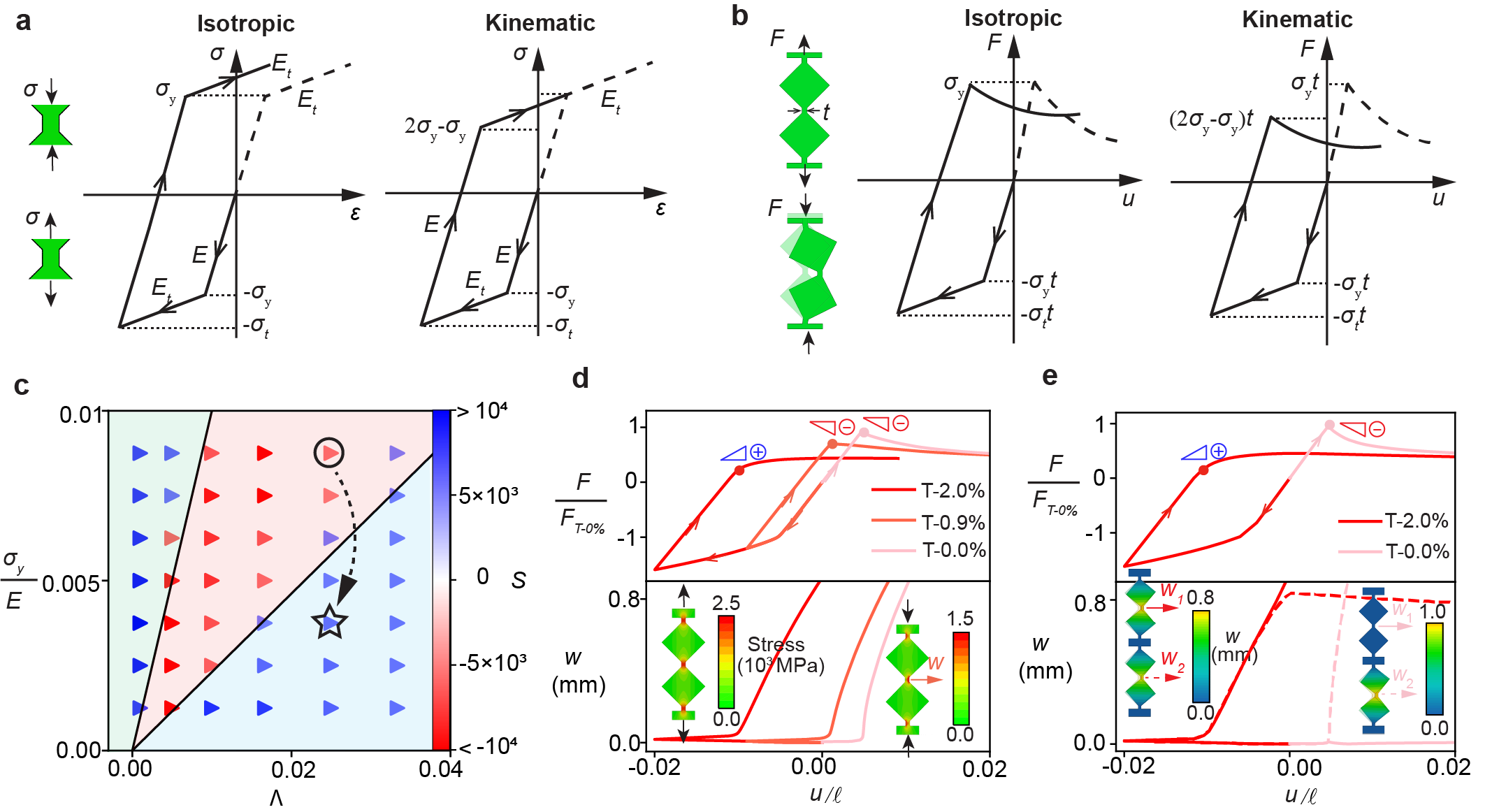}
\caption{\textbf{Yield buckling tuned by loading history.}
\textbf{a and b,} Isotropic and kinematic work hardening rules in (a) the elastoplastic material and (b) the yield buckling unit cell. 
\textbf{c,} Stiffness $S$ at the buckling moment in the regimes of elastic, plastic, and yield buckling tuned by the yield stress $\sigma_{\text{y}}/E$ and aspect ratio $\Lambda$~\cite{Liu2024harness}. 
\textbf{d,} Force $F_{\text{T}}/F_{\text{T}-0\%}$ and lateral deflection $W$ vs. stroke $u/\ell$ of a unit cell under three different stretching and compressing loading histories. The force $F$ is normalized by the critical buckling force of the unit cell without stretching $F_{\text{T}-0\%}$. The insets are the von Mises stress state of a unit cell under $0.9\%$ of stretching (left) and at the onset of buckling (right).
\textbf{e,} Force $F/F_{\text{T}-0\%}$ and lateral deflection $W$ vs. stroke $u/\ell$ of the two-layer structure under two different loading histories.The insets are the deflection states of the two unit cells at the onset of buckling under two different loading histories.
}
\label{fig:7}
\end{figure*}

Now we demonstrate simplified work hardening rules with a bilinear plastic model in a uniaxial loading condition (Fig.~\ref{fig:7}a) and apply them to the unit cell in the regime of yield buckling (Fig.~\ref{fig:7}b). The idea is that we first stretch the unit cell and make the ligaments yield and strengthen at a stress $\sigma_{t}$ ($\sigma_{t}>\sigma_{\text{y}}$) and then compress it until buckling happens. For the isotropic model, the unit cell will not be affected by the stretching behavior and will buckle at the same yield stress $\sigma_{\text{y}}$ (Fig.~\ref{fig:7}b solid line) as the unit cell without stretching (Fig.~\ref{fig:7}b dashed line). In contrast, the yield buckling stress of the unit cell can reduce to $2\sigma_{\text{y}}-\sigma_{t}$ 
after stretching at a stress $\sigma_{t}$ for the linear kinematic hardening model (Fig.~\ref{fig:7}b right). We numerically prove this strategy in a unit cell located the the regime of yield buckling (Fig.~\ref{fig:7}c circle). We choose three different stretching strains 0\%, 0.9\%, and 2.0\% before compression and find the  Bauschinger effect can effectively reduce the buckling load in yield buckling but does not affect the elastic and plastic buckling. 
Interestingly, the Bauschinger effect can also switch the stiffness of the unit cell at the onset of buckling from negative to positive (Fig.~\ref{fig:7}d) where the unit cell sits in the regime of plastic buckling (Fig.~\ref{fig:7}c star) when the yield load is lower than the reduced modulus load of the unit cell. 

The buckling load reduction and stiffness transition at the onset of buckling due to the Bauschinger effect can be used to design multifunctional metamaterials with different loading histories. We numerically prove this idea by applying a bilinear plastic model with the kinematic work hardening rule to a structure that consists of two lumped unit cells with the same geometry. In the regime of yield buckling, this structure first shows sequential buckling under compression since the negative stiffness at the onset of buckling (see the pink force and deflection vs displacement curves in Fig.~\ref{fig:7}e). We then stretch the structure to a stroke of 2.0$\%$ before compressing. The stiffness of the unit cells transfers from negative to position and the structure shows simultaneous buckling (Fig.~\ref{fig:7}e red curves).

\section{Discussion and conclusion}

We have demonstrated that the yielding area (ligament thickness) and the yield criterion (ligament orientation) of the unit cell can be used to tune the yield load and yield buckling sequence in metamaterials. The difference between these two strategies is that the buckling sequence tuned by the yield criterion becomes less robust when the number of buckling sequences increases. The yielding area is more robust in tuning sequential buckling over a large range of sequences and can lead to a tunable effective hardening. 
Using yield area and criterion in combination can significantly broaden the design space for a larger buckling sequence in large-scale sequential metamaterials.
Last, we explored the multiple buckling sequences in metamaterials tuned with the loading history and the Bauschinger effect. One thing worth mentioning is that when the unit cell is slender enough and made from a elastoplastic material, it can also buckle before the ligaments yield. In this case, the Bauschinger effect can be used to change the buckling of the unit cell from the regime of elastic buckling to the regime of yield buckling load and even to the regime of plastic buckling. The Bauschinger effect as an interesting and important phenomenon has been observed in many metallic materials, but it is rarely used in the design of materials or metamaterials in practice. In this work, we add the Bauschinger effect as a design tool in tuning the buckling behaviors in metamaterials and prove such concepts with a numerical model. However, the use of the Bauschinger effect is still challenging in the design of metamaterials. One reason is that the difference between stretching and compressing yield stress caused by the Baushinger effect is very small and is usually neglected in practice. Another reason is the fatigue of the elastoplastic materials under cyclic load. How to amplify and exploit such Bauschinger effect remains hence a fascinating future research direction. More generally, the next question is how does fatigue affect the buckling behavior of metamaterials? Can fatigue of the material also be used as a design tool to tune the buckling behavior of metamaterials with loading history? 

Our work enriches the toolbox of plasticity in the design of controllable yield bucking and sequential metamaterials and we anticipate such controllable yield bucking can lead to distinctive applications in controllable shock absorption~\cite{yuan2023butterfly}, mechanical self-assembly~\cite{van2020kirigami, meng2022deployable}, multifunctional soft robotics~\cite{hwang2022shape, wang2023robotic}, and physical learning in materials~\cite{stern2020supervised}.

%In conclusion, we further explored yielding area, yield criteria, and loading history with the Buashinger effect in plasticity as design tools in tuning yield buckling and the yield buckling sequence of sequential metamaterials. We found that the buckling sequence in the regime of yield buckling is dominated by the yielding area (ligament thickness) of the unit cells. Instead, the buckling sequence is controlled by the aspect ratio of the unit cells in the regimes of elastic and plastic buckling. We then introduced general yield criteria in tuning the yield load of the buckling unit cell by changing the orientation angle of the ligaments. We experimentally prove such two strategies in different metamaterial architectures over a large number of controllable buckling sequences. Lastly, we introduced the Buashinger effect from the kinematic work hardening rule into metamaterials with yield buckling and achieved multiple buckling sequences in a single geometry with loading history.

\emph{Acknowledgments.} 
We thank Caroline Kopecz-Muller and Yuan Zhou for insightful discussions and suggestions, Daan Giesen, Sven Koot for technical assistance. We acknowledge funding from the European Research Council under grant agreement 852587 and the Netherlands Organisation for Scientific Research under grant agreement NWO TTW 17883.
All the codes and data supporting this study are available on the public repository  \url{https://doi.org/0.5281/zenodo.13945234}.
\bibliographystyle{unsrt}
\bibliography{refs.bib}
\clearpage

%%%%%%%%%%
%%%%%%%%%%
%%%%%%%%%%
%%%%%%%%%%
%%%%%%%%%%
\setcounter{equation}{0}
\renewcommand{\theequation}{A\arabic{equation}}%
\setcounter{figure}{0}
\renewcommand{\figurename}{Extended Data Fig.}

\textbf{\huge Appendix}
\vspace{12pt}
\section{Free-standing sequential metamaterials}

\begin{figure*}[h!]
\centering
\includegraphics[width=1.0\linewidth]{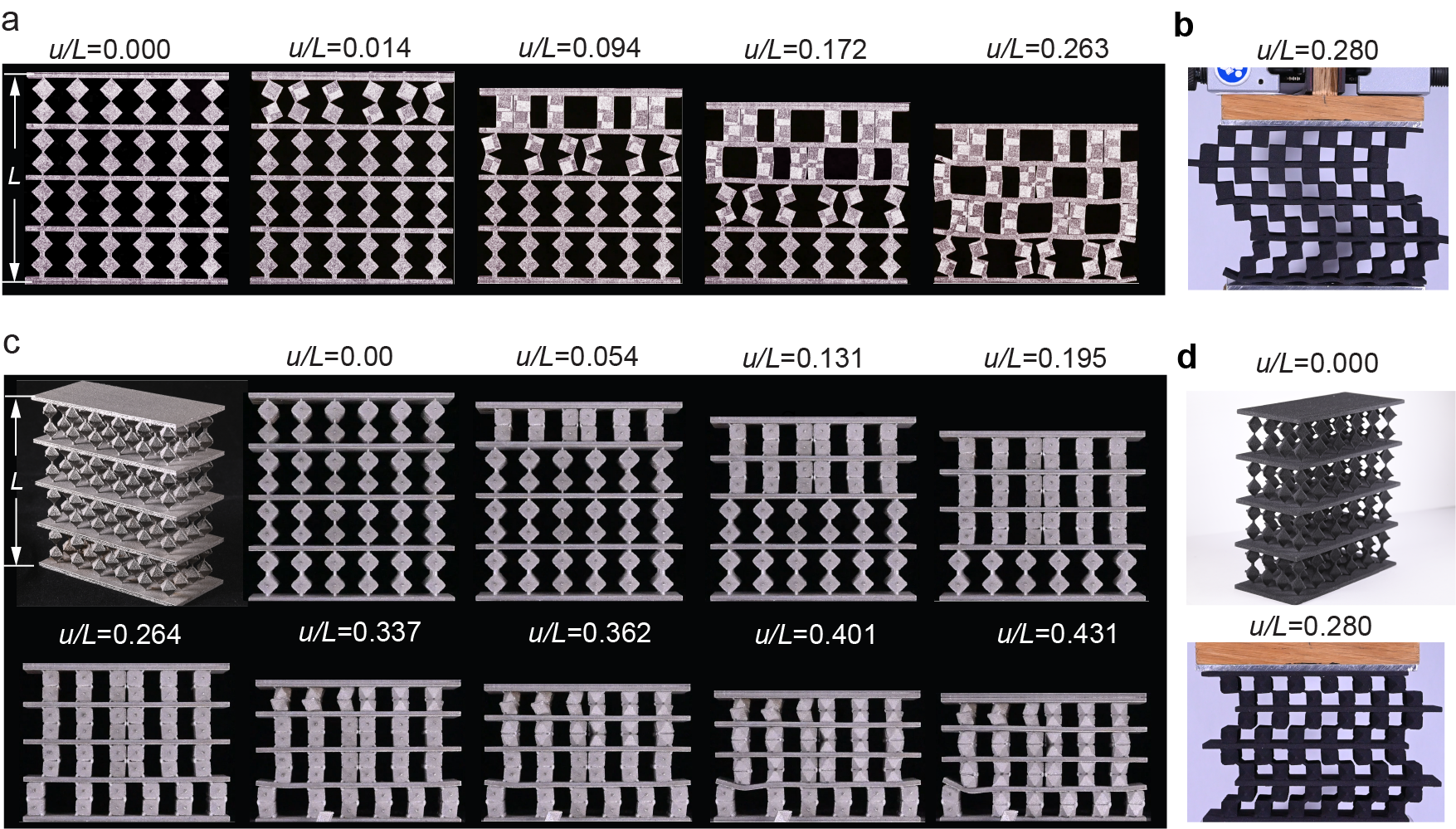}
\caption{\textbf{Sequential yield buckling in metamaterials without side constraints.} Four-layer metamaterials of ligament thickness ratio $t_{2}/t_{1}=t_{3}/t_{2}=t_{4}/t_{3}=1.8$ between each buckling layer and without the need of constraints at the sides to prevent shearing under uniaxial compression and the compressive speed is 0.2 mm s$^{-1}$ unless indicated otherwise. \textbf{a and b,} Snapshots of metamaterials with two-square buckling unit cells at different compressive strokes $u/L$, (a) a metamaterial made from 316L stainless steel and (b) a metamaterial made from elastic printing material. \textbf{c and d,} Snapshots of metamaterials with two-octahedron buckling unit cells at different compressive strokes $u/L$, (a) a metamaterial made from 316L stainless steel and (b) a metamaterial made from elastic printing material.}
\label{fig:8}
\end{figure*}

In the Main Text, we have shown that the yield area (ligament thickness) and yield criterion (ligament orientation) can be used to tune the sequence of yield buckling in metamaterials. Compared to the ligament orientation, varying the ligament thickness in metamaterials shows a more robust buckling sequence with a larger difference in the buckling load between buckling layers. Here we further demonstrate that a larger difference in the ligament thickness between the buckling layers can overcome the global shearing and have a robust sequential buckling in the metamaterials (Fig.~\ref{fig:8}a and c) without side constraints. As a comparison, we also conduct compression experiments in metamaterials with the same geometry printed from elastic material which have a global shearing mode (Fig.~\ref{fig:8}b and d).

% \bibliographystylemethod{unsrt}
% \bibliographymethod{refs.bib}

\vspace{12pt}

%%%%%%%%%%%%%%extended figures%%%%%%%%%%%%

\end{document}